\documentclass[aip,rsi,amsmath,amssymb,reprint]{revtex4-2}

\usepackage{graphicx}
\usepackage{color}
\usepackage[dvipsnames]{xcolor}
\usepackage{float}
\usepackage{dcolumn}
\usepackage{bm}
\usepackage{url}

\usepackage[utf8]{inputenc}
\usepackage[T1]{fontenc}
\usepackage{mathptmx}

\begin{document}

\preprint{AIP/123-QED}

\title[Petrovi{\'c} \textit{et al.}]{A perpendicular field electromagnet with a 250mm access bore}

\author{A.P. Petrovi{\'c}}
 \thanks{Authors to whom correspondence should be addressed:\\ appetrovic@ntu.edu.sg, christos@ntu.edu.sg}
\author{B.H.M. Smit}%
 \altaffiliation[Also at ]{Department of Applied Physics, TU Eindhoven, The Netherlands.}
\author{K.L. Fong}
 \altaffiliation[Also at ]{Department of Physics, The Hong Kong University of Science and Technology, Hong Kong.}
\author{B. Satywali}
\author{X.Y. Tee}
\author{C. Panagopoulos}
 \thanks{Authors to whom correspondence should be addressed:\\ appetrovic@ntu.edu.sg, christos@ntu.edu.sg}
\affiliation{Division of Physics and Applied Physics, School of Physical and Mathematical Sciences, Nanyang Technological University, 21 Nanyang Link, Singapore 637371}

\date{\today}

\begin{abstract}
We present a laboratory electromagnet capable of generating magnetic fields up to $\pm$~0.48~T, specifically designed as a perpendicular flux source for thin film samples in an ambient environment.  The magnet features a 250~mm diameter clear access bore above the sample plane, thus offering compatibility with a wide variety of experimental apparatus.  Despite its generous size, the magnet thermally dissipates less than 1~kW at maximum field.  A shaped ferromagnetic core is used to amplify and homogenize the field $\mathbf{B}$, leading to an estimated uniformity of $\pm$1.5~mT ($\lesssim0.3$\%) in $\left|\mathbf{B}\right|$ within a 28~mm$^2$ zone at maximum field.  The sample stage is thermally regulated and isolated from the magnet, enabling temperature control with $\pm$5~mK precision even at elevated magnetic fields.  
\end{abstract}

\maketitle

\section{\label{Intro}Introduction\protect\\}
Many experiments in modern condensed matter physics necessitate the application of a magnetic field perpendicular to a thin film sample or device.  Magnetic films with perpendicular anisotropy (PMA) are of considerable importance for spintronic applications~\cite{Piramanayagam2007,Ikeda2010} and provide an excellent example of this requirement.  In such materials, the spin-orbit interaction causes a preferential orientation of the electron spins perpendicular to the film plane~\cite{Hellman2017}, in contrast to the in-plane alignment which would be expected from magnetostatic and exchange effects alone.  Combining PMA with a spin-canting interaction such as geometric frustration or the Dzyaloshinskii-Moriya interaction can induce the formation of chiral spin textures such as skyrmions~\cite{Moreau-Luchaire2016,Soumyanarayanan2017}, with significant potential for data storage, logic or neuromorphic computing~\cite{Fert2013a,Song2020a}.  

Technological applications of PMA films and/or chiral spin textures will involve the controlled reversal of out-of-plane magnetic moments.  Typical examples include flipping the spin polarisation in PMA tunnel junctions~\cite{Igarashi2017}, modifying the domain structure in PMA nanodots~\cite{Moutafis2007}, strips~\cite{Lavrijsen2010} or films~\cite{Talapatra2016,Shepley2018}, and nucleating/translating/erasing individual magnetic skyrmions.  Ultimately, such magnetic switching may be achieved within multilayer devices by applying spin torques from an out-of-plane current~\cite{Mangin2006}.  However, if direct access to the exposed magnetic layer is necessary during the research and development phase - as is the case for any imaging or local probe experiment - then an external perpendicular flux source becomes essential to manipulate the film magnetization.  

Integrating experimental scanning or imaging hardware with a perpendicular magnetic field is technically challenging.  Room-temperature laboratory electromagnets generally use a split pair of coils, each surrounding a conical pole made from a high permeability material (e.g. pure iron).  These poles focus the magnetic flux into the experimental area within the gap between them.  Following this principle, a coil can be wound around a C-shaped core or yoke, and the sample-holder placed in the gap between the ends of the yoke~\cite{Williams1982}.  The coil can alternatively be replaced by a flux-coupled permanent magnet array~\cite{Proksch1995,Gomez1996,Mohanty2005,Harroun2006}, with the advantage that resistive heating is eliminated.  Unfortunately, none of these approaches can be used to generate a perpendicular field for room-temperature apparatus such as a force microscope or nanoprobe station, since the upper pole of the yoke would completely cover the sample.  A limited degree of optical access in perpendicular fields has been achieved using hollow magnetic poles~\cite{Oldenbourg1986} or a ``magic mangle'' configuration of ferromagnetic cylinders~\cite{Cugat1994}, but both these approaches still severely obstruct the sample surface.  

Previous room-temperature imaging studies of PMA films have circumnavigated this problem by using arrays of permanent magnets underneath the sample to generate an out-of-plane field~\cite{Soumyanarayanan2017,Duong2019}.  Although effective, this approach has two disadvantages: firstly, the magnetic field strength cannot be continuously varied, thus precluding detailed studies of the magnetization evolution and reversal processes.  Secondly, the field from a typical disc-shaped permanent magnet can exhibit substantial ($>\pm$10\%) inhomogeneity over a typical $\sim5\times5$~mm substrate, leading to unwanted flux gradients at macroscopic lengthscales.  Commercial systems based on movable permanent magnet arrays have been recently developed for force microscopy: these offer a higher flux homogeneity, but only generate perpendicular flux densities up to $\pm$0.12~T~\cite{Asylum} or $\pm$0.02~T~\cite{NTEGRA}.  Such field strengths fall well below the 0.3-0.4~T required to polarise many chiral magnetic films.  Another commercial electromagnet provides perpendicular fields of up to 0.95~T for magneto-optical Kerr effect microscopy, using an annular flux yoke above the sample~\cite{AEvico1}.  However, the optical path to the sample passes through a narrow hole in the yoke, thus preventing access to the surface using bulky scanning/probing hardware.  An alternative electromagnet from the same supplier without any flux yoke can only provide fields up to 0.1~T~\cite{BEvico2}, illustrating the constraints imposed by the requirement of unrestricted sample access.

\begin{figure*}[htbp]
\includegraphics[trim=0cm 0cm 0cm 0cm, clip=true, width=2\columnwidth]{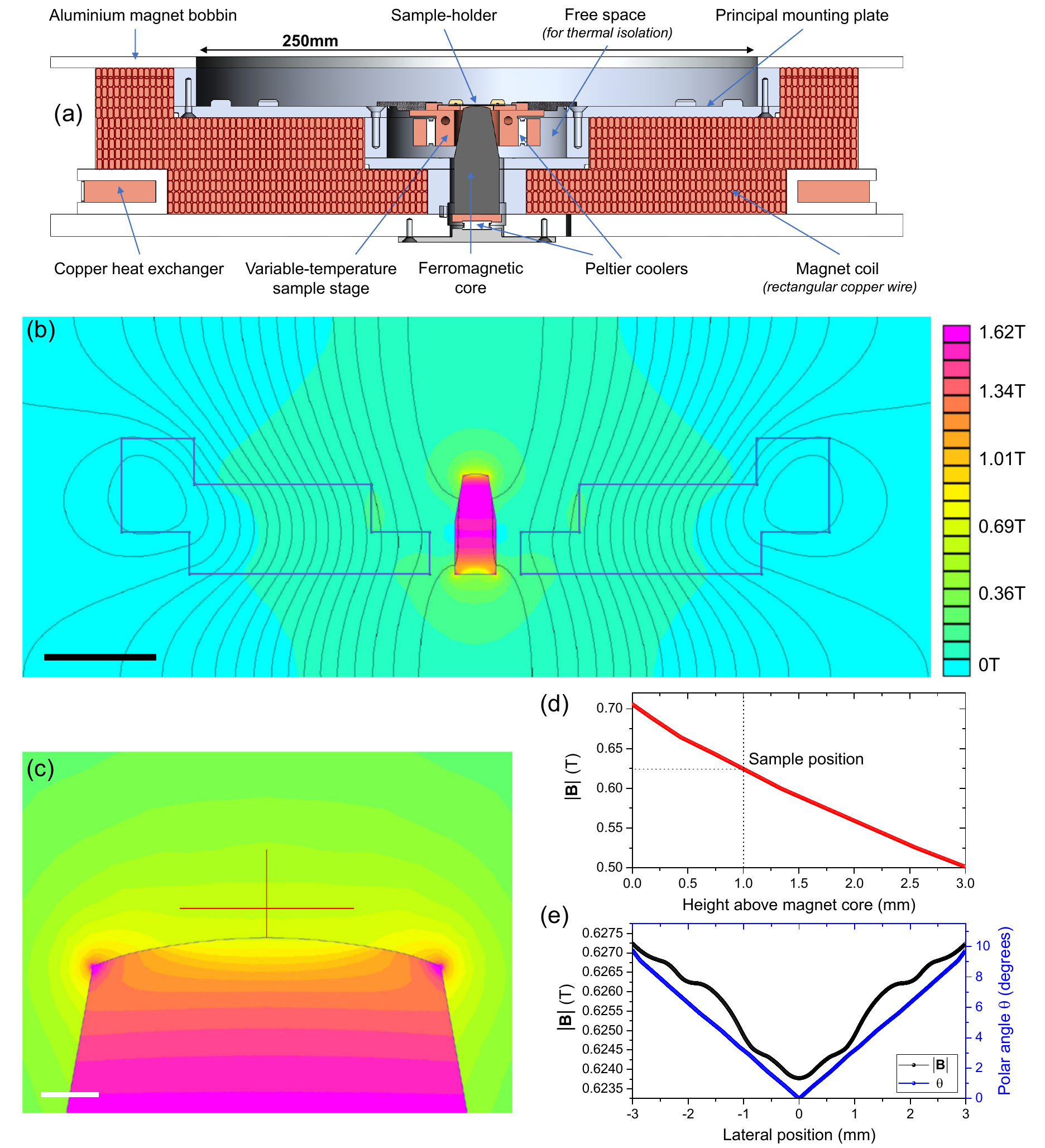}
\caption{\label{Fig1} (a) Cross-sectional slice through the centre of the magnet design.  The 250~mm diameter zone above the sample-holder constitutes the only spatial constraint on any experimental hardware used to analyse the thin film under test.  The large area of the principal mounting plate (to which the thermally-isolated sample-holder is fixed) provides adequate space for installing a nanoprobe station or microscope, thus minimising any thermal drift between probe and sample.  (b) Simulated flux density generated by the electromagnet driven by current $I_{max}=$~28~A.  Thin black lines are flux lines, i.e. contours of the magnetic vector potential $\mathbf{A}$ multiplied by $2\pi r$ (where $r$ is the distance from the magnet central axis).  The region occupied by the coil is delineated by blue lines and the iron core is visible as the high flux density zone at the centre of the magnet.  The black scale bar measures 50~mm.  (c) Zoom view of the top surface of the iron core, using the same colour scheme to illustrate the flux density as in (b).  The 6~mm horizontal red line lies in the sample plane 1~mm above the core, while the 3~mm vertical red line lies along the central axis.  The white scale bar measures 2~mm.  (d) Decay of the flux density moving up the vertical red line in (c): at the sample height, the field is decreasing at 0.064~mT\,$\mu$m$^{-1}$.  (e) Homogeneity of the absolute flux density $\left|\mathbf{B}\right|$ and the polar angle $\theta$ subtended by the local magnetic field within the sample plane (horizontal red line in (c)). }
\end{figure*}

These continuity, homogeneity and field strength problems could all be eliminated by placing the entire sample and apparatus into the bore of a superconducting solenoid.  However, the requirement for cryogenic cooling adds considerable complexity to the measurement process (as well as expense during setup and operation).  Replacing the superconducting coil with a resistive solenoid operating at room temperature yields magnetic fields which are too weak to be useful: for example, a copper coil with an internal bore $\sim$~50~mm dissipating several kW in resistive heating only generates a field $\sim$~0.05T.  Moreover, high-field solenoids typically have narrow bore diameters < 100~mm, necessitating custom miniaturized experimental hardware to fit inside the magnet.

Here we present a new electromagnet design which overcomes all these challenges to provide a strong, homogeneous and continuously-variable magnetic field perpendicular to the plane of a thin film sample.  Above the sample plane, the magnet bore is sufficiently large (250~mm diameter) to ensure compatibility with many commercial imaging or analysis tools.  The magnet assembly process does not require any specialised tools or techniques and can hence be adapted for use in any laboratory.  Most importantly, the magnet dissipates less than 1~kW at maximum field (0.48~T) and has a sufficiently low inductance to enable pulsing from zero to maximum field at timescales of the order of 1~s.  These factors both minimise the heat load on the sample stage during magnet operation, providing an exceptionally thermally stable environment.

\section{Magnet design}

Our initial approach employs a large (0.34~m outer diameter) solenoid coil with a ``stepped'' internal diameter and a high purity (>99.8\%) iron core.  A cross-section of this configuration is shown in Fig.~\ref{Fig1}(a), illustrating the core position directly ($\leq1$~mm) underneath the sample.  The coil and core geometries have been designed to maximize the perpendicular field strength and homogeneity within a 6~mm diameter zone in the sample plane using FEMM finite-element magnetostatic analysis software~\cite{FEMM}.  In particular, the three ``steps'' in the coil cross-section allow us to minimize the total resistance (and hence power dissipation) of the coil, while incorporating a thermally-isolated sample stage and optimizing the field perpendicularity at the sample plane.  A 28~A 36~V bipolar power supply~\cite{AKepcoBOP} was then selected to match the calculated resistance (1.2$\Omega$) and inductance (0.11~H) of the coil.  

The simulated field from the solenoid (driven at the maximum current, $I_{max}=28$~A) is illustrated in Fig.~\ref{Fig1}(b).  Here we have assumed an optimistic 90\% coil-packing efficiency, corresponding to 689 turns of a 2.52~mm$^2$ cross-section copper wire.  At the designated sample height (1~mm above the iron core), the vertical flux density $B$ reaches 0.624~T on the core central axis and the magnet is dissipating 959~W in heat due to resistive losses in the copper wire.  A zoom view of the field in the sample zone directly above the iron core is shown in Fig.~\ref{Fig1}(c), together with line cuts of the flux density vertically above the core centre and laterally at 1~mm above the core (the designated sample position).  At this height, the vertical field is falling at 0.064~mT\,$\mu$m$^{-1}$ (Fig.~\ref{Fig1}(d)).  Given that the typical thin film heterostructures which we wish to study have thicknesses below 100~nm, this decay is negligible.  Homogeneity within the $xy$ plane is more difficult to achieve: although the absolute flux density varies by less than $\pm$0.3\% within a 6~mm diameter circle, the field orientation deviates increasingly from the perpendicular, reaching a maximum polar angle of 9.5$^\circ$ at the edge of the circle (Fig.~\ref{Fig1}(e)).  The homogeneity of the field amplitude and orientation are controlled by the curvature of the top surface of the iron core: flattening this surface would improve the orientational homogeneity, at the expense of the amplitude homogeneity.  We emphasize that within the central 40~$\mu$m~$\times$~40~$\mu$m region - the maximal scan area of a typical scanning probe microscope - our core design generates an estimated flux density which is homogeneous to within 0.005\%, with a maximum polar offset $\leq0.1^\circ$.    

\begin{figure}[tbp]
\includegraphics[trim=0cm 0cm 0cm 0cm, clip=true, width=1\columnwidth]{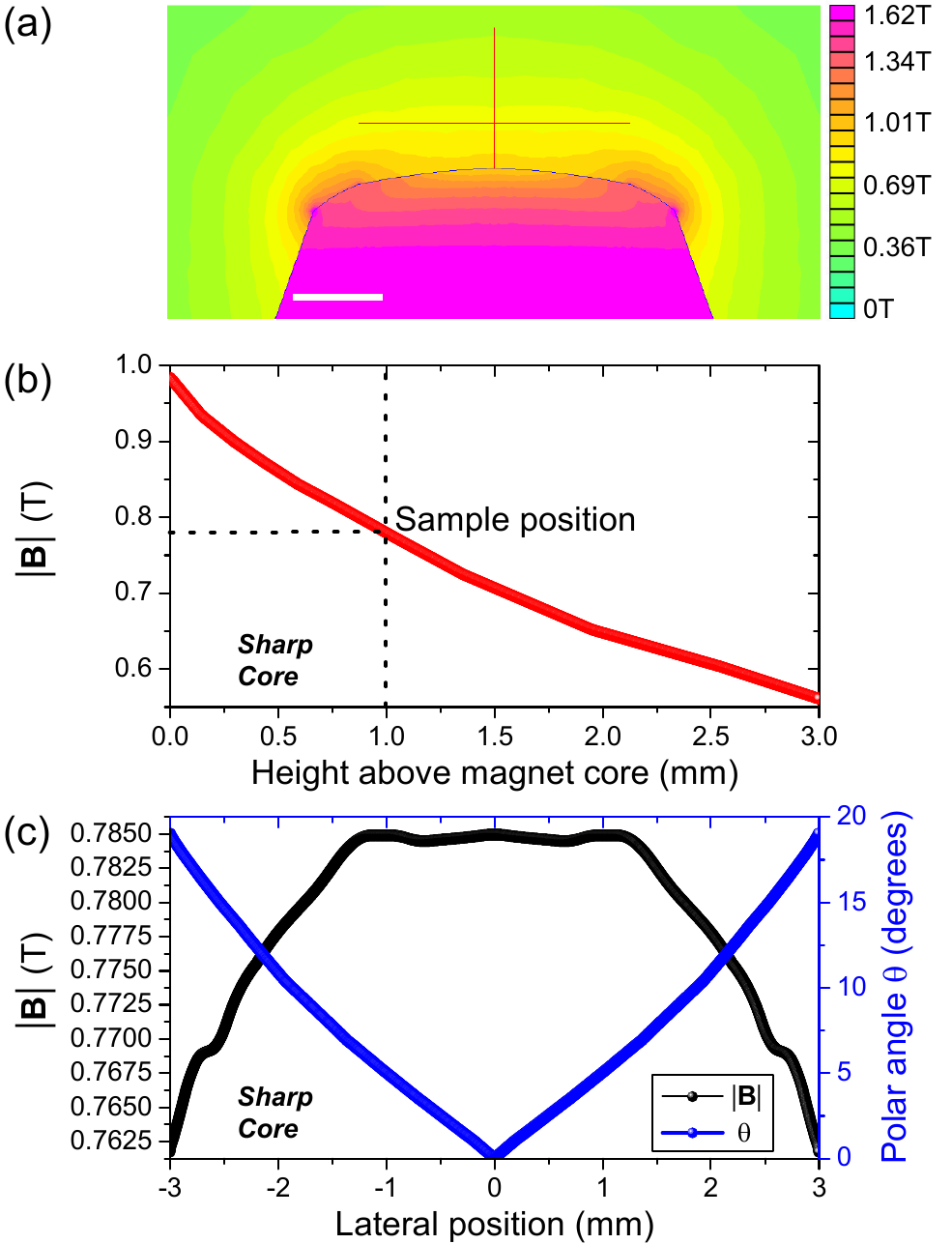}
\caption{\label{Fig2} Influence of changing the core geometry on the achievable field strength and homogeneity.  (a) Simulated flux density above a broader, sharply-tapering iron core at $I_{max}=28$~A.  The base and tip diameters of the core are 23.6~mm and 8~mm respectively, compared with 20~mm and 12~mm for the simulations shown in Fig.~\ref{Fig1}(b-e).  The white scale bar measures 2~mm.  Apart from the core geometry, all other simulation parameters are identical to those used in Fig.~\ref{Fig1}.  (b) Decay of the flux density moving up the core axis (vertical red line in (a)): at the sample height, the field is decreasing at 0.157~mT\,$\mu$m$^{-1}$.  (c) Homogeneity of the absolute flux density $\left|\mathbf{B}\right|$ and the polar angle $\theta$ subtended by the field within the sample plane (horizontal red line in (a)).}
\end{figure}

Changing the geometry of the core offers a simple route to tuning both the magnitude and homogeneity of the applied field.  In Fig.~\ref{Fig2}(a) we show a simulation of the field profile created above an iron core with a more sharply-tapered tip, which serves to concentrate the flux into the sample.  The maximum perpendicular flux density at the sample plane increases to 0.785~T (Fig.~\ref{Fig2}(b)) and we are able to optimize the tip curvature to obtain a highly homogeneous field ($\pm$0.05\%) in a central 2.5~mm diameter circle (Fig.~\ref{Fig2}(c)).  However, the reduced core radius causes the field to decay outside this zone, resulting in a larger inhomogeneity of $\pm$1.5\% within a 6mm diameter circle.  Moreover, the orientational homogeneity deteriorates, with the field developing a polar angle of 19$^\circ$ at radius 3~mm.  These data demonstrate that optimal core design for a particular experiment must balance the required field magnitude against the sample dimensions, as well as the tolerance of the intended measurement to deviations in field orientation.  

\section{Magnet construction}
The magnet bobbin is assembled from a set of four custom-machined plates and three cylinders, all made in 6061-T6 aluminium.  This material was chosen for four reasons: it is non-magnetic, has a high thermal conductivity (to minimise thermal gradients across the coil), low mass and is easy to machine.  The plates were fastened to the cylinders using countersunk non-magnetic SS-316L screws and threadlocking compound.  Helicoil thread inserts in the aluminium were avoided due to their ferromagnetic nature.  Prior to winding the coil, all internal surfaces of the bobbin were lined with 50~$\mu$m Kapton film, which was bonded to the aluminium using a thin layer of thermally-conductive epoxy~\cite{Masterbond}: this eliminated the possibility of short-circuiting the coil by scratching the wire insulation during assembly.  

\begin{figure}[tbp]
\includegraphics[trim=0cm 0cm 0cm 0cm, clip=true, width=1\columnwidth]{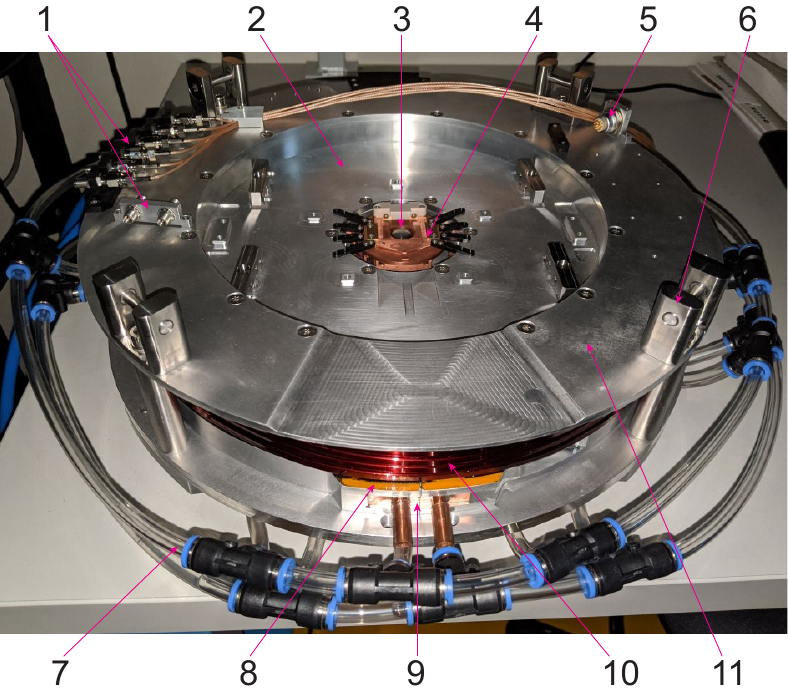}
\caption{\label{Fig3} Photograph of our as-built magnet and sample stage.  The dielectric substrate has been removed from the sample-holder (exposing the ferromagnetic core below) and the electrical cables leading to the sample stage have been disconnected for clarity.  Labels indicate: 1) coaxial cable junctions for \textit{in situ} high frequency experiments; 2) principal mounting plate (featuring multiple threaded holes for installing experimental hardware); 3) ferromagnetic core; 4) variable-temperature sample stage; 5) multi-pin connector for heaters, Peltier coolers and thermometers; 6) one of four attachment points (allowing the entire magnet to be suspended for vibration isolation during sensitive experiments); 7) chilled water supply lines; 8) Kapton insulation protecting the magnet bobbin from electrical shorts; 9) one of nine copper heat exchangers; 10) outer diameter of the magnet coil; 11) aluminium magnet bobbin.}
\end{figure}

\begin{figure}[bp]
\includegraphics[clip=true, width=0.99\columnwidth]{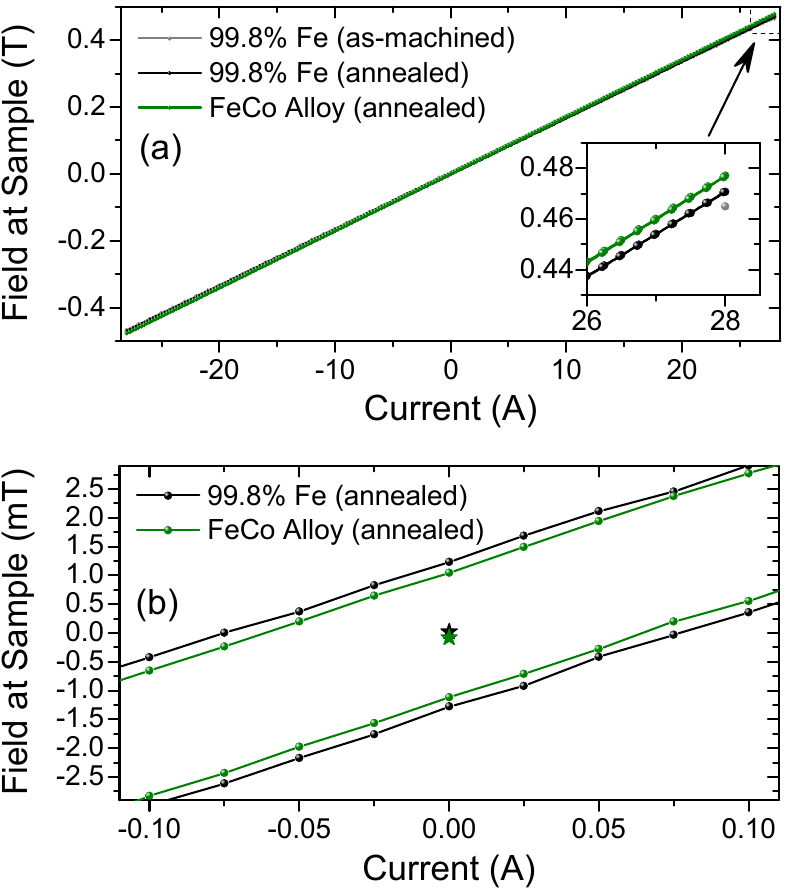}
\caption{\label{Fig4} (a) Field-current calibration performed at the sample stage of the electromagnet.  Data from three different ferromagnetic cores are shown: a 99.8\% pure iron core immediately after fabrication, the same iron core following a 10-hour annealing process, and an annealed Fe$_{49}$Co$_{49}$V$_2$ alloy.  The inset highlights the small (1\%) improvement offered by the FeCo alloy compared with the annealed iron.  (b) Magnetic hysteresis loops for annealed iron and FeCo cores (black and green data-points, respectively) illustrating the remanent fields in each core.  The stars at zero current indicate the measured fields achieved after performing an automated demagnetization loop.}
\end{figure}

To maximise the current density in the coil, we used 1.45$\times$4.37~mm rectangular copper wire~\cite{Superior} with 110~$\mu$m insulation: this was the largest cross-sectional area which could easily conform to the 13~mm minimum bend radius required by our bobbin.  The coil was hand-wound using a simple home-made jig which allowed the magnet bobbin and $\sim$~100~kg wire reel to rotate independently while keeping the wire parallel to both bobbins.  From our original estimate of 689 total coils (Fig.~1(b)), we anticipated winding a total of 85 wire layers on the bobbin.  Upon completion of each layer, a high-viscosity magnet resin~\cite{Elantas} was applied by brush.  This particular resin was selected for its rapid gelling time at room temperature (4 hours).  At the end of the winding process, the magnet bobbin contained 622 coils: this reduction from the predicted value (689) can be attributed to excess resin application.  After winding, the magnet was heated to 60$^\circ$C in air to complete the resin curing.  Electrical connections to the magnet were made using two 9~mm diameter copper welding cables, terminated by gold-plated copper blocks clamped onto the ends of the magnet wire.  Each power terminal is isolated from the magnet bobbin using a Delrin plate.  An annotated photograph of the as-built magnet is shown in Fig.~\ref{Fig3}: its total mass is 41~kg.  

\section{Magnetic field calibration}

\begin{figure*}[htbp]
\includegraphics[trim=0.1cm 7.2cm 0.1cm 2.2cm, clip=true, width=1.9\columnwidth]{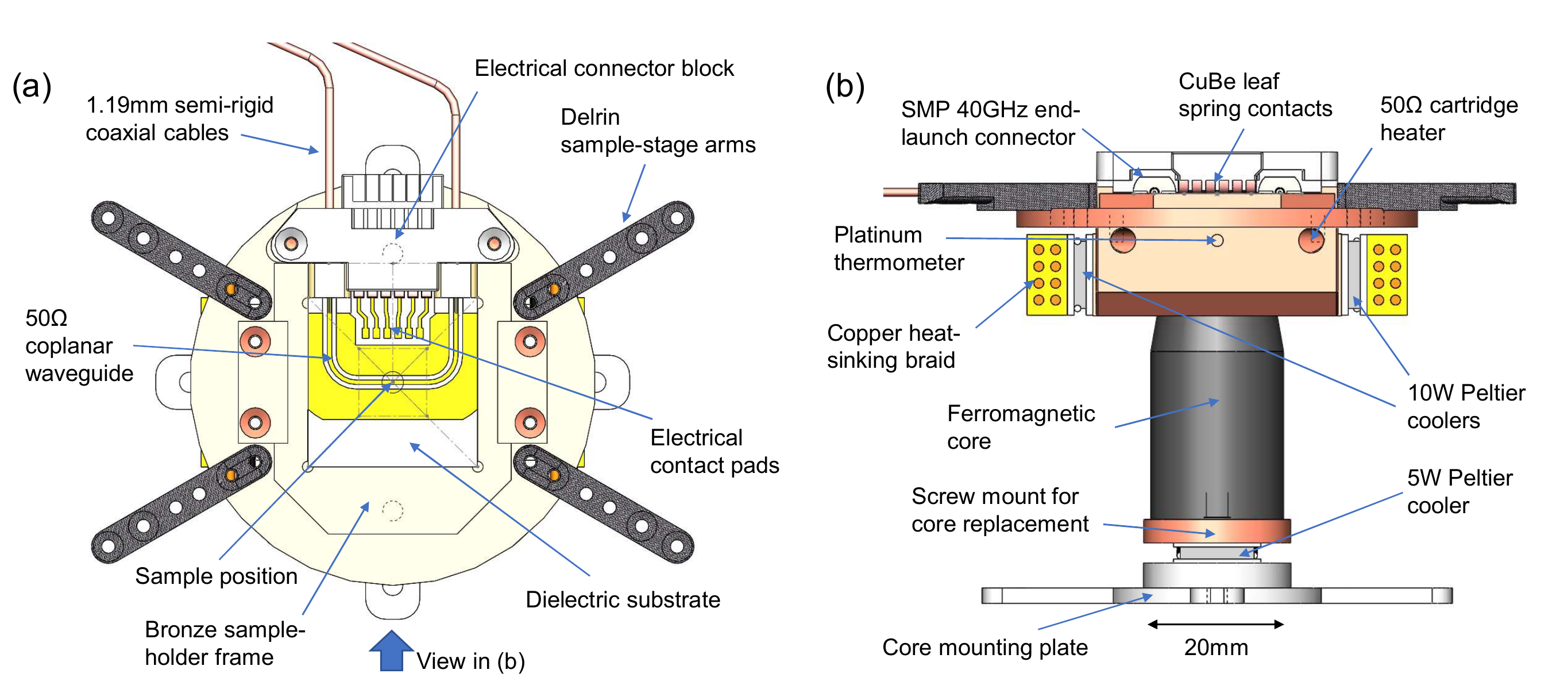}
\caption{\label{Fig5} Variable-temperature sample stage viewed from (a) above and (b) the side, along the direction indicated by the blue arrow in (a).  The Delrin arms provide a stiff, low thermal conductivity interface to the principal mounting plate on the magnet body.  Samples are loaded using a removable sample-holder plate, which slides horizontally into position parallel to the viewing arrow in (a).  Each sample-holder incorporates two SMP end-launch coaxial connectors which mate with female SMP jacks in the PEEK connector block, thus firmly securing the sample-holder.  Up to six further electrical contacts to the substrate can be made using a series of 0.1~mm copper beryllium leaf springs, positioned between the two SMP jacks.  Beneath the copper thermal block, the ferromagnetic core screws into a threaded mount, thus facilitating replacement of the core for tuning the field magnitude and homogeneity.}
\end{figure*} 

The magnet is operated via a LabVIEW program allowing real-time control and monitoring of the field as well as the sample and magnet temperatures.  This software also facilitates calibration of the magnet, i.e. determination of the current-field conversion factor.  The magnetic field is measured using a factory-calibrated Hall sensor~\cite{Hall} mounted on a sample-holder at the same height as our experimental samples.  The results are shown in Fig.~\ref{Fig4}(a): using a freshly-machined 99.8\% iron core with identical geometry to the simulations in Fig.~\ref{Fig1}, a maximum field of $\pm0.465$~T at $I_{max}=28$~A is achieved, which is lower than the expected 0.624~T.  An obvious source of error is the reduced coil count (622 vs. 689) in our as-built magnet.  However, correcting the number of coils and rerunning our simulations yields a maximum field of 0.566~T, which is still 20\% higher than our measured field.  This suggests that the offset originates either from the magnetic properties of the core, or from the radial current density within the magnet windings.  

We first consider the contribution of the core.  Our simulation in Fig.~\ref{Fig1}(b) uses the reference permeability for pure iron $\mu_r =$~14872 included in the FEMM package, but it is well known that this value can be strongly suppressed by structural defects induced during the machining process and/or the presence of impurities.  We therefore performed a vacuum annealing at 820$^\circ$C for 10 hours to improve the microstructure of our iron core, but this only increased the maximum field by 1\% to 0.471~T.  Replacing the iron core with a high-induction FeCo alloy~\cite{Vacoflux} provided another 1\% increase to 0.478~T.  We therefore believe that the principal limitation in our magnet performance originates from a reduced coil density close to the centre of the bobbin.  The innermost coils of the magnet were the most awkward to wind tightly due to their small radius $r$, yet according to the Biot-Savart law $B = \mu_0I/2r$ provide the largest contribution to the flux density at the centre of the magnet.  As illustrated in Fig.~\ref{Fig2}, a simple low-cost route to improve the magnet performance would therefore be to increase the ferromagnetic core diameter (up to a maximum of 23.6~mm imposed by the bobbin internal diameter) to capture more flux, and sharpen the tip of the core (at the expense of the lateral field homogeneity at the sample).  However, we emphasize that the $\pm$0.48~T achievable with our existing core geometry is already sufficient to polarise and reverse the majority of thin film heterostructures based on elemental magnets (e.g. PtCo, TaCoFeB, IrFeCoPt, etc.).  Moreover, this field is considerably larger than the $\sim$~0.1T achievable by commercial magnets suitable for unconstrained imaging in perpendicular fields~\cite{Asylum,NTEGRA,BEvico2}.  

The $B(I)$ curves in Fig.~\ref{Fig4}(a) appear perfectly linear, with a gradient of 0.0171~TA$^{-1}$ for the FeCo core.  However, using a single linear factor to convert from magnet current to sample field becomes inaccurate at very low fields.  Despite our deliberate selection of low remanence materials for the magnet core, some hysteresis-induced error in the magnetic field experienced by the sample is inevitable.  We quantify and correct for this error by calibrating the magnetic field during a full $B(I)$ hysteresis loop between $\pm$28~A.  Figure~\ref{Fig4}(b) highlights the low-current region of the hysteresis loops acquired with our annealed iron and FeCo cores: a remanent field of $1.25\pm0.05$~mT persists at $I=0$ in the iron core after ramping down from maximum field.  The FeCo alloy is marginally magnetically softer than pure iron, with a remanent field of $1.05\pm0.05$~mT.  Measurements in true zero field can be achieved by performing an automated demagnetization sequence, in which a series of $n$ currents $I_i= I_{max}\times(-A)^i$ is applied to the solenoid.  Here $A\sim0.8$ is an adjustable parameter, $I_1\sim20$~A, $i=1:n$ and the sequence terminates when $I_n$ falls below a user-configurable threshold $\sim$~3~mA ($\approx$~0.05~mT).  After completing this cycle, the core is fully demagnetized and the residual field at the sample is less than 0.1~mT.  

\section{Variable-temperature sample stage}

A major limitation of any laboratory electromagnet is the heat dissipated by the resistive coil.  Since the magnetic properties of thin film materials can vary strongly with temperature (especially their anisotropy), it is essential to manage this heat flow and hence maintain a known, constant temperature at the sample/device under analysis.  Our solution to this problem is shown in Fig.~\ref{Fig5}: we place the sample on a massive (0.2~kg) copper block suspended from the principal mounting plate by stiff Delrin arms, thus thermally decoupling the sample environment from the magnet body.  The copper block houses two 50~$\Omega$ cartridge heaters (wired in parallel for redundancy) and a platinum resistive thermometer~\cite{ALakeshorePt}, which are connected to a PID temperature controller~\cite{BLakeshore}. On each side of the block, a 10.4~W Peltier cooler~\cite{Peltier} is mounted using thermally conductive silver epoxy~\cite{AEpotek1}.  The hot side of each Peltier chip is heatsunk to the magnet body using flexible copper braid.  An additional 5.1~W Peltier cooler~\cite{Peltier} is positioned under the ferromagnetic core, below a threaded mount which facilitates core replacement.  The maximum current rating for each Peltier chip is 2.2~A.  All three coolers are wired in series and powered by a programmable 0-2.5~A current source~\cite{BKepcoPTR}, which in turn is manually adjusted using an analog output from the temperature controller.  A further platinum thermometer is mounted on the magnet body for monitoring purposes.  

The centre of the copper thermal block is hollow, allowing it to encircle the ferromagnetic core.  Our sample-holders are composed of a 25$\times$25$\times$0.25mm dielectric substrate~\cite{Rogers}, with a ground plane deposited on the underside.  This substrate can be patterned with electrical contacts or waveguides to allow \textit{in situ} excitations and measurements from dc to 40~GHz.  Samples are affixed directly on top of the substrate, using dilute GE varnish for electrical isolation if appropriate.  For ease of handling and insertion, the substrate is glued into a U-shaped bronze frame using thermally conductive epoxy~\cite{BEpotek2}.  This assembly slides into position on top of the sample stage, guided by two copper rails to mate with a set of six electrical and two microwave connectors housed in an insulating PEEK frame.  After inserting the sample-holder into this connector block, the separation between the substrate ground plane and the shaped core below is 200~$\mu$m.  The sample is hence electrically and thermally isolated from the magnet bobbin.  

\begin{figure}[htbp]
\includegraphics[clip=true, width=0.99\columnwidth]{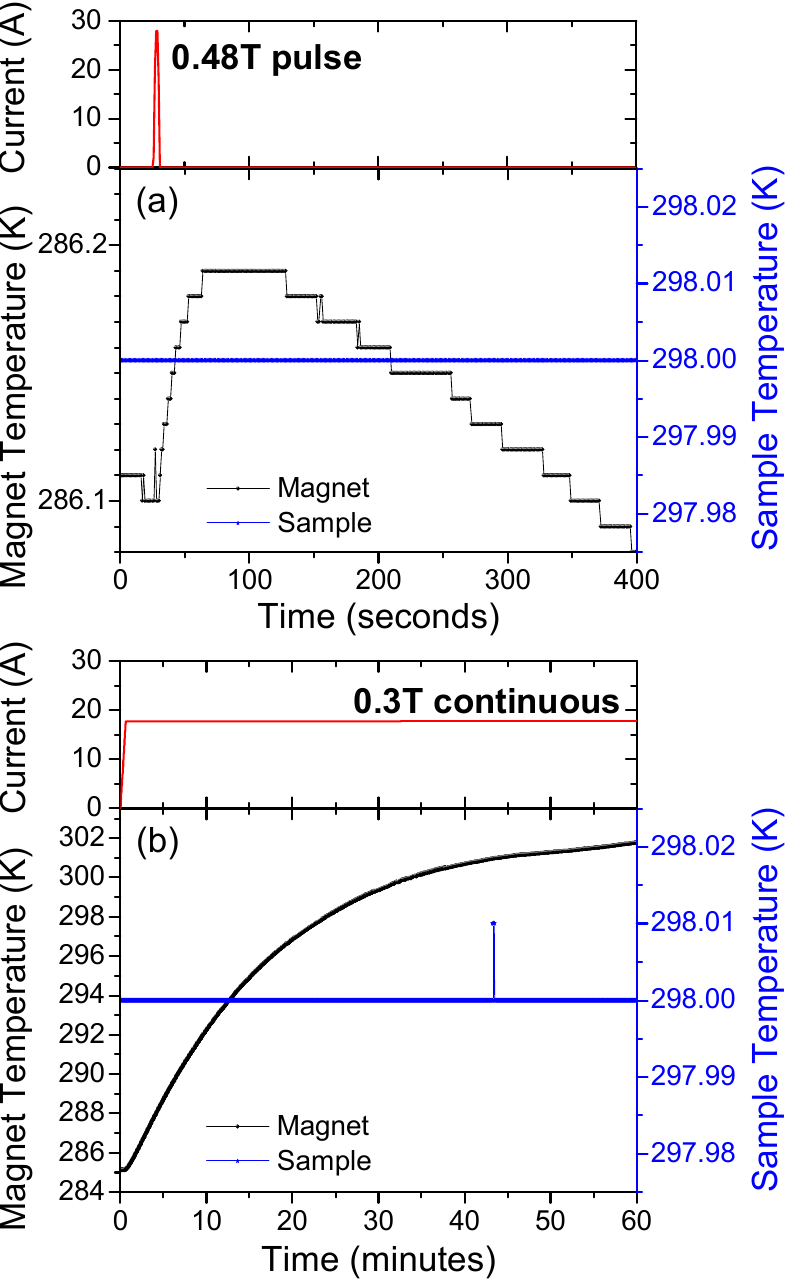}
\caption{\label{Fig6} (a) Thermal stability of the magnet and sample stage during magnet pulsing to maximum field.  The slow drift to lower temperatures is due to the cooling from the 10$^\circ$C chilled water supply.  (b) Long-term (1 hour) thermal stability of the magnet and sample at a constant field of 0.3~T.  The data acquisition frequency is 1~Hz; the sample temperature only fluctuated above the $\pm$5~mK resolution limit of the temperature controller for one second during the entire test period.}
\end{figure}

In addition to regulating the temperature of the sample stage, it is also important to prevent the magnet from overheating.  During operation, heat is removed from the magnet by nine water-cooled copper heat exchangers~\cite{HExch}.  Five of these are fixed to the underside of the magnet bobbin using brass screws and thermal paste, while the other four are attached to aluminium wedges embedded within the coil.  All heat exchangers can be replaced if they develop leaks during use.  A 0.6~kW recirculating chiller~\cite{Julabo} pumps water through all nine heat exchangers in parallel, at a maximum rate of 15~l~min$^{-1}$.  We find that a temperature setpoint of 10$^\circ$C provides the optimal balance between maximal cooling power at high currents and minimal condensation on the magnet bobbin in zero field.  

While current is flowing through the magnet, a thermal gradient develops between the upper and lower surfaces of the bobbin.  This gradient is not detrimental to the thermal stability of our sample stage, which only contacts the magnet bobbin at the principal mounting plate.  However, the magnet cooling efficiency could certainly be improved by mounting additional heat exchangers on the top surface of the bobbin.  We have chosen not to implement this step in our magnet, to maximise the free space available for installing experimental hardware.  

\section{Magnet thermal stability}

We define two conditions to assess the thermal performance of our electromagnet: short-term temperature stability in response to rapidly-changing fields, and long-term stability in constant fields.  

An important feature of our control software is a ``Pulse'' function which ramps the magnet current momentarily to a (high) setpoint, then returns immediately to a lower constant field.  This is useful for quickly polarizing a magnetic film while avoiding any thermal instability in the magnet or sample.  Our power supply has a high bandwidth (>800~Hz) and hence the ramp-rate limiting factor is its 36~V maximum output voltage, since the potential across the coil has both resistive $IR$ and inductive $LdI/dT$ components during magnet charging/discharging.  We find that a minimum rate of 20~As$^{-1}$ ($\sim0.4$~Ts$^{-1}$) can be achieved even close to maximum field, which is consistent with the measured magnet resistance 1.22~$\Omega$ and calculated inductance (0.083~H for 622 coils).  

We demonstrate the thermal stability of the magnet during a 2~second pulse to maximum field in Fig.~\ref{Fig6}(a).  Before applying the pulse, the magnet is cooled to $\sim$~286~K by the 10$^\circ$C chilled water circuit.  The sample stage is stabilised at 298~K by regulating the heater output from the temperature controller while passing a constant 0.7~A current through the Peltier coolers.  Following the pulse, the magnet body temperature only rises by $0.1$~K over a 30 second period.  However, the sample stage temperature displays no discernible rise above the 5~mK resolution threshold of our temperature controller.  The magnet relaxes back to its original temperature within 6 minutes.

The long-term thermal stability (which is crucial for extended imaging experiments at high fields) is illustrated in Fig.~\ref{Fig6}(b).  We pass a constant 17.6~A current through the magnet (corresponding to a 0.3~T field at the sample), while continuously monitoring the magnet and sample temperatures for 1 hour.  Although the magnet temperature rises by 16~K during this period, the sample stage is maintained at 298~K with the same $\pm$5~mK stability.  Extended testing at maximum field indicated that the sample stage temperature can be stabilised at 300~K until the magnet body temperature exceeds $\approx$~318~K.  At such high temperatures, the principal limiting factor in our sample stage design is the heatsinking of the two Peltier coolers on the copper sample block.  If we attempt to recover thermal stability by increasing the Peltier current (and hence raising the cooling power), the waste heat at the hot side of the Peltier chips cannot dissipate sufficiently quickly into the magnet body and the sample stage is thermally overloaded.  

The thermal decoupling between sample stage and magnet body also enables experiments to be performed above room temperature.  In its present configuration, our apparatus is limited to operation below 90$^\circ$C, since the solder joints in the Peltier coolers risk failure through thermal fatigue above this temperature.  The Delrin arms on the sample stage are not expected to deform until at least 120$^\circ$C, while the insulation materials used to assemble the magnet are rated for use up to 180$^\circ$C.  We therefore anticipate that by substituting the Delrin with a carbon fiber matrix (or a similar heat-tolerant, low thermal conductivity material), and replacing the Peltier coolers with versions adapted for high temperature use, measurements could be performed up to at least 180$^\circ$C.  At this temperature, we estimate that the magnet body would experience a heat load $\approx2$~W from the sample stage, which is negligible in comparison to the power dissipation within the coil.

\section{Conclusions}

We have designed and built a $\pm$0.48~T laboratory electromagnet which is optimised for studying thin film materials in a perpendicular field.  The wide 250~mm bore of the magnet above the sample plane provides unlimited access to the sample surface even for large apparatus such as scanning probe microscopes.  The sample temperature can be regulated with $\pm$5~mK precision independently of changes in the magnet current or temperature, thus providing a thermally stable measurement environment.  Finally, the ease with which the ferromagnetic core can be machined and replaced promises to facilitate a variety of experiments at even higher field strengths and/or in spatially-varying flux profiles.  

\section{Data Availability}
The data which support the findings of this study are available from the corresponding authors upon reasonable request.

\begin{acknowledgments}
We acknowledge support from the Singapore National Research Foundation (NRF) NRF-Investigatorship (No. NRFNRFI2015-04), Singapore MOE Academic Research Fund Tier 3 Grant MOE2018-T3-1-002 and Tier 1 Grant M4012006. We are grateful to Alex Rajan and Chandrabose Rajagopal from Ebenezer Exel Engineering for technical assistance.  
\end{acknowledgments}


\begin{thebibliography}{41}%
\makeatletter
\providecommand \@ifxundefined [1]{%
 \@ifx{#1\undefined}
}%
\providecommand \@ifnum [1]{%
 \ifnum #1\expandafter \@firstoftwo
 \else \expandafter \@secondoftwo
 \fi
}%
\providecommand \@ifx [1]{%
 \ifx #1\expandafter \@firstoftwo
 \else \expandafter \@secondoftwo
 \fi
}%
\providecommand \natexlab [1]{#1}%
\providecommand \enquote  [1]{``#1''}%
\providecommand \bibnamefont  [1]{#1}%
\providecommand \bibfnamefont [1]{#1}%
\providecommand \citenamefont [1]{#1}%
\providecommand \href@noop [0]{\@secondoftwo}%
\providecommand \href [0]{\begingroup \@sanitize@url \@href}%
\providecommand \@href[1]{\@@startlink{#1}\@@href}%
\providecommand \@@href[1]{\endgroup#1\@@endlink}%
\providecommand \@sanitize@url [0]{\catcode `\\12\catcode `\$12\catcode
  `\&12\catcode `\#12\catcode `\^12\catcode `\_12\catcode `\%12\relax}%
\providecommand \@@startlink[1]{}%
\providecommand \@@endlink[0]{}%
\providecommand \url  [0]{\begingroup\@sanitize@url \@url }%
\providecommand \@url [1]{\endgroup\@href {#1}{\urlprefix }}%
\providecommand \urlprefix  [0]{URL }%
\providecommand \Eprint [0]{\href }%
\providecommand \doibase [0]{https://doi.org/}%
\providecommand \selectlanguage [0]{\@gobble}%
\providecommand \bibinfo  [0]{\@secondoftwo}%
\providecommand \bibfield  [0]{\@secondoftwo}%
\providecommand \translation [1]{[#1]}%
\providecommand \BibitemOpen [0]{}%
\providecommand \bibitemStop [0]{}%
\providecommand \bibitemNoStop [0]{.\EOS\space}%
\providecommand \EOS [0]{\spacefactor3000\relax}%
\providecommand \BibitemShut  [1]{\csname bibitem#1\endcsname}%
\let\auto@bib@innerbib\@empty
\bibitem [{\citenamefont {Piramanayagam}(2007)}]{Piramanayagam2007}%
  \BibitemOpen
  \bibfield  {author} {\bibinfo {author} {\bibfnamefont {S.~N.}\ \bibnamefont
  {Piramanayagam}},\ }\bibfield  {title} {\enquote {\bibinfo {title}
  {{Perpendicular recording media for hard disk drives}},}\ }\href@noop {}
  {\bibfield  {journal} {\bibinfo  {journal} {Journal of Applied Physics}\
  }\textbf {\bibinfo {volume} {102}},\ \bibinfo {pages} {011301} (\bibinfo
  {year} {2007})}\BibitemShut {NoStop}%
\bibitem [{\citenamefont {Ikeda}\ \emph {et~al.}(2010)\citenamefont {Ikeda},
  \citenamefont {Miura}, \citenamefont {Yamamoto}, \citenamefont {Mizunuma},
  \citenamefont {Gan}, \citenamefont {Endo}, \citenamefont {Kanai},
  \citenamefont {Hayakawa}, \citenamefont {Matsukura},\ and\ \citenamefont
  {Ohno}}]{Ikeda2010}%
  \BibitemOpen
  \bibfield  {author} {\bibinfo {author} {\bibfnamefont {S.}~\bibnamefont
  {Ikeda}}, \bibinfo {author} {\bibfnamefont {K.}~\bibnamefont {Miura}},
  \bibinfo {author} {\bibfnamefont {H.}~\bibnamefont {Yamamoto}}, \bibinfo
  {author} {\bibfnamefont {K.}~\bibnamefont {Mizunuma}}, \bibinfo {author}
  {\bibfnamefont {H.~D.}\ \bibnamefont {Gan}}, \bibinfo {author} {\bibfnamefont
  {M.}~\bibnamefont {Endo}}, \bibinfo {author} {\bibfnamefont {S.}~\bibnamefont
  {Kanai}}, \bibinfo {author} {\bibfnamefont {J.}~\bibnamefont {Hayakawa}},
  \bibinfo {author} {\bibfnamefont {F.}~\bibnamefont {Matsukura}},\ and\
  \bibinfo {author} {\bibfnamefont {H.}~\bibnamefont {Ohno}},\ }\bibfield
  {title} {\enquote {\bibinfo {title} {{A perpendicular-anisotropy CoFeB-MgO
  magnetic tunnel junction}},}\ }\href@noop {} {\bibfield  {journal} {\bibinfo
  {journal} {Nature Materials}\ }\textbf {\bibinfo {volume} {9}},\ \bibinfo
  {pages} {721} (\bibinfo {year} {2010})}\BibitemShut {NoStop}%
\bibitem [{\citenamefont {Hellman}\ \emph {et~al.}(2017)\citenamefont
  {Hellman}, \citenamefont {Hoffmann}, \citenamefont {Tserkovnyak},
  \citenamefont {Beach}, \citenamefont {Fullerton}, \citenamefont {Leighton},
  \citenamefont {MacDonald}, \citenamefont {Ralph}, \citenamefont {Arena},
  \citenamefont {D{\"{u}}rr}, \citenamefont {Fischer}, \citenamefont
  {Grollier}, \citenamefont {Heremans}, \citenamefont {Jungwirth},
  \citenamefont {Kimel}, \citenamefont {Koopmans}, \citenamefont {Krivorotov},
  \citenamefont {May}, \citenamefont {Petford-Long}, \citenamefont
  {Rondinelli}, \citenamefont {Samarth}, \citenamefont {Schuller},
  \citenamefont {Slavin}, \citenamefont {Stiles}, \citenamefont {Tchernyshyov},
  \citenamefont {Thiaville},\ and\ \citenamefont {Zink}}]{Hellman2017}%
  \BibitemOpen
  \bibfield  {author} {\bibinfo {author} {\bibfnamefont {F.}~\bibnamefont
  {Hellman}}, \bibinfo {author} {\bibfnamefont {A.}~\bibnamefont {Hoffmann}},
  \bibinfo {author} {\bibfnamefont {Y.}~\bibnamefont {Tserkovnyak}}, \bibinfo
  {author} {\bibfnamefont {G.~S.~D.}\ \bibnamefont {Beach}}, \bibinfo {author}
  {\bibfnamefont {E.~E.}\ \bibnamefont {Fullerton}}, \bibinfo {author}
  {\bibfnamefont {C.}~\bibnamefont {Leighton}}, \bibinfo {author}
  {\bibfnamefont {A.~H.}\ \bibnamefont {MacDonald}}, \bibinfo {author}
  {\bibfnamefont {D.~C.}\ \bibnamefont {Ralph}}, \bibinfo {author}
  {\bibfnamefont {D.~A.}\ \bibnamefont {Arena}}, \bibinfo {author}
  {\bibfnamefont {H.~A.}\ \bibnamefont {D{\"{u}}rr}}, \bibinfo {author}
  {\bibfnamefont {P.}~\bibnamefont {Fischer}}, \bibinfo {author} {\bibfnamefont
  {J.}~\bibnamefont {Grollier}}, \bibinfo {author} {\bibfnamefont {J.~P.}\
  \bibnamefont {Heremans}}, \bibinfo {author} {\bibfnamefont {T.}~\bibnamefont
  {Jungwirth}}, \bibinfo {author} {\bibfnamefont {A.~V.}\ \bibnamefont
  {Kimel}}, \bibinfo {author} {\bibfnamefont {B.}~\bibnamefont {Koopmans}},
  \bibinfo {author} {\bibfnamefont {I.~N.}\ \bibnamefont {Krivorotov}},
  \bibinfo {author} {\bibfnamefont {S.~J.}\ \bibnamefont {May}}, \bibinfo
  {author} {\bibfnamefont {A.~K.}\ \bibnamefont {Petford-Long}}, \bibinfo
  {author} {\bibfnamefont {J.~M.}\ \bibnamefont {Rondinelli}}, \bibinfo
  {author} {\bibfnamefont {N.}~\bibnamefont {Samarth}}, \bibinfo {author}
  {\bibfnamefont {I.~K.}\ \bibnamefont {Schuller}}, \bibinfo {author}
  {\bibfnamefont {A.~N.}\ \bibnamefont {Slavin}}, \bibinfo {author}
  {\bibfnamefont {M.~D.}\ \bibnamefont {Stiles}}, \bibinfo {author}
  {\bibfnamefont {O.}~\bibnamefont {Tchernyshyov}}, \bibinfo {author}
  {\bibfnamefont {A.}~\bibnamefont {Thiaville}},\ and\ \bibinfo {author}
  {\bibfnamefont {B.~L.}\ \bibnamefont {Zink}},\ }\bibfield  {title} {\enquote
  {\bibinfo {title} {{Interface-Induced Phenomena in Magnetism}},}\ }\href
  {http://www.ncbi.nlm.nih.gov/pubmed/28890576{\%}0Ahttp://www.pubmedcentral.nih.gov/articlerender.fcgi?artid=PMC5587142}
  {\bibfield  {journal} {\bibinfo  {journal} {Reviews of Modern Physics}\
  }\textbf {\bibinfo {volume} {89}},\ \bibinfo {pages} {025006} (\bibinfo
  {year} {2017})}\BibitemShut {NoStop}%
\bibitem [{\citenamefont {Moreau-Luchaire}\ \emph {et~al.}(2016)\citenamefont
  {Moreau-Luchaire}, \citenamefont {Moutafis}, \citenamefont {Reyren},
  \citenamefont {Sampaio}, \citenamefont {Vaz}, \citenamefont {{Van Horne}},
  \citenamefont {Bouzehouane}, \citenamefont {Garcia}, \citenamefont
  {Deranlot}, \citenamefont {Warnicke}, \citenamefont {Wohlh{\"{u}}ter},
  \citenamefont {George}, \citenamefont {Weigand}, \citenamefont {Raabe},
  \citenamefont {Cros},\ and\ \citenamefont {Fert}}]{Moreau-Luchaire2016}%
  \BibitemOpen
  \bibfield  {author} {\bibinfo {author} {\bibfnamefont {C.}~\bibnamefont
  {Moreau-Luchaire}}, \bibinfo {author} {\bibfnamefont {C.}~\bibnamefont
  {Moutafis}}, \bibinfo {author} {\bibfnamefont {N.}~\bibnamefont {Reyren}},
  \bibinfo {author} {\bibfnamefont {J.}~\bibnamefont {Sampaio}}, \bibinfo
  {author} {\bibfnamefont {C.~A.~F.}\ \bibnamefont {Vaz}}, \bibinfo {author}
  {\bibfnamefont {N.}~\bibnamefont {{Van Horne}}}, \bibinfo {author}
  {\bibfnamefont {K.}~\bibnamefont {Bouzehouane}}, \bibinfo {author}
  {\bibfnamefont {K.}~\bibnamefont {Garcia}}, \bibinfo {author} {\bibfnamefont
  {C.}~\bibnamefont {Deranlot}}, \bibinfo {author} {\bibfnamefont
  {P.}~\bibnamefont {Warnicke}}, \bibinfo {author} {\bibfnamefont
  {P.}~\bibnamefont {Wohlh{\"{u}}ter}}, \bibinfo {author} {\bibfnamefont
  {J.-M.}\ \bibnamefont {George}}, \bibinfo {author} {\bibfnamefont
  {M.}~\bibnamefont {Weigand}}, \bibinfo {author} {\bibfnamefont
  {J.}~\bibnamefont {Raabe}}, \bibinfo {author} {\bibfnamefont
  {V.}~\bibnamefont {Cros}},\ and\ \bibinfo {author} {\bibfnamefont
  {A.}~\bibnamefont {Fert}},\ }\bibfield  {title} {\enquote {\bibinfo {title}
  {{Additive interfacial chiral interaction in multilayers for stabilization of
  small individual skyrmions at room temperature}},}\ }\href@noop {} {\bibfield
   {journal} {\bibinfo  {journal} {Nature Nanotechnology}\ }\textbf {\bibinfo
  {volume} {11}},\ \bibinfo {pages} {444} (\bibinfo {year} {2016})}\BibitemShut
  {NoStop}%
\bibitem [{\citenamefont {Soumyanarayanan}\ \emph {et~al.}(2017)\citenamefont
  {Soumyanarayanan}, \citenamefont {Raju}, \citenamefont {{Gonzalez Oyarce}},
  \citenamefont {Tan}, \citenamefont {Im}, \citenamefont {Petrovi{\'{c}}},
  \citenamefont {Ho}, \citenamefont {Khoo}, \citenamefont {Tran}, \citenamefont
  {Gan}, \citenamefont {Ernult},\ and\ \citenamefont
  {Panagopoulos}}]{Soumyanarayanan2017}%
  \BibitemOpen
  \bibfield  {author} {\bibinfo {author} {\bibfnamefont {A.}~\bibnamefont
  {Soumyanarayanan}}, \bibinfo {author} {\bibfnamefont {M.}~\bibnamefont
  {Raju}}, \bibinfo {author} {\bibfnamefont {A.~L.}\ \bibnamefont {{Gonzalez
  Oyarce}}}, \bibinfo {author} {\bibfnamefont {A.~K.~C.}\ \bibnamefont {Tan}},
  \bibinfo {author} {\bibfnamefont {M.~Y.}\ \bibnamefont {Im}}, \bibinfo
  {author} {\bibfnamefont {A.~P.}\ \bibnamefont {Petrovi{\'{c}}}}, \bibinfo
  {author} {\bibfnamefont {P.}~\bibnamefont {Ho}}, \bibinfo {author}
  {\bibfnamefont {K.~H.}\ \bibnamefont {Khoo}}, \bibinfo {author}
  {\bibfnamefont {M.}~\bibnamefont {Tran}}, \bibinfo {author} {\bibfnamefont
  {C.~K.}\ \bibnamefont {Gan}}, \bibinfo {author} {\bibfnamefont
  {F.}~\bibnamefont {Ernult}},\ and\ \bibinfo {author} {\bibfnamefont
  {C.}~\bibnamefont {Panagopoulos}},\ }\bibfield  {title} {\enquote {\bibinfo
  {title} {{Tunable room-temperature magnetic skyrmions in Ir/Fe/Co/Pt
  multilayers}},}\ }\href {https://doi.org/10.1038/NMAT4934} {\bibfield
  {journal} {\bibinfo  {journal} {Nature Materials}\ }\textbf {\bibinfo
  {volume} {16}},\ \bibinfo {pages} {898--904} (\bibinfo {year}
  {2017})}\BibitemShut {NoStop}%
\bibitem [{\citenamefont {Fert}, \citenamefont {Cros},\ and\ \citenamefont
  {Sampaio}(2013)}]{Fert2013a}%
  \BibitemOpen
  \bibfield  {author} {\bibinfo {author} {\bibfnamefont {A.}~\bibnamefont
  {Fert}}, \bibinfo {author} {\bibfnamefont {V.}~\bibnamefont {Cros}},\ and\
  \bibinfo {author} {\bibfnamefont {J.}~\bibnamefont {Sampaio}},\ }\bibfield
  {title} {\enquote {\bibinfo {title} {{Skyrmions on the track}},}\ }\href@noop
  {} {\bibfield  {journal} {\bibinfo  {journal} {Nature Nanotechnology}\
  }\textbf {\bibinfo {volume} {8}},\ \bibinfo {pages} {152--156} (\bibinfo
  {year} {2013})}\BibitemShut {NoStop}%
\bibitem [{\citenamefont {Song}\ \emph {et~al.}(2020)\citenamefont {Song},
  \citenamefont {Jeong}, \citenamefont {Pan}, \citenamefont {Zhang},
  \citenamefont {Xia}, \citenamefont {Cha}, \citenamefont {Park}, \citenamefont
  {Kim}, \citenamefont {Finizio}, \citenamefont {Raabe}, \citenamefont {Chang},
  \citenamefont {Zhou}, \citenamefont {Zhao}, \citenamefont {Kang},
  \citenamefont {Ju},\ and\ \citenamefont {Woo}}]{Song2020a}%
  \BibitemOpen
  \bibfield  {author} {\bibinfo {author} {\bibfnamefont {K.~M.}\ \bibnamefont
  {Song}}, \bibinfo {author} {\bibfnamefont {J.~S.}\ \bibnamefont {Jeong}},
  \bibinfo {author} {\bibfnamefont {B.}~\bibnamefont {Pan}}, \bibinfo {author}
  {\bibfnamefont {X.}~\bibnamefont {Zhang}}, \bibinfo {author} {\bibfnamefont
  {J.}~\bibnamefont {Xia}}, \bibinfo {author} {\bibfnamefont {S.}~\bibnamefont
  {Cha}}, \bibinfo {author} {\bibfnamefont {T.~E.}\ \bibnamefont {Park}},
  \bibinfo {author} {\bibfnamefont {K.}~\bibnamefont {Kim}}, \bibinfo {author}
  {\bibfnamefont {S.}~\bibnamefont {Finizio}}, \bibinfo {author} {\bibfnamefont
  {J.}~\bibnamefont {Raabe}}, \bibinfo {author} {\bibfnamefont
  {J.}~\bibnamefont {Chang}}, \bibinfo {author} {\bibfnamefont
  {Y.}~\bibnamefont {Zhou}}, \bibinfo {author} {\bibfnamefont {W.}~\bibnamefont
  {Zhao}}, \bibinfo {author} {\bibfnamefont {W.}~\bibnamefont {Kang}}, \bibinfo
  {author} {\bibfnamefont {H.}~\bibnamefont {Ju}},\ and\ \bibinfo {author}
  {\bibfnamefont {S.}~\bibnamefont {Woo}},\ }\bibfield  {title} {\enquote
  {\bibinfo {title} {{Skyrmion-based artificial synapses for neuromorphic
  computing}},}\ }\href@noop {} {\bibfield  {journal} {\bibinfo  {journal}
  {Nature Electronics}\ }\textbf {\bibinfo {volume} {3}},\ \bibinfo {pages}
  {148--155} (\bibinfo {year} {2020})}\BibitemShut {NoStop}%
\bibitem [{\citenamefont {Igarashi}\ \emph {et~al.}(2017)\citenamefont
  {Igarashi}, \citenamefont {Llandro}, \citenamefont {Sato}, \citenamefont
  {Matsukura},\ and\ \citenamefont {Ohno}}]{Igarashi2017}%
  \BibitemOpen
  \bibfield  {author} {\bibinfo {author} {\bibfnamefont {J.}~\bibnamefont
  {Igarashi}}, \bibinfo {author} {\bibfnamefont {J.}~\bibnamefont {Llandro}},
  \bibinfo {author} {\bibfnamefont {H.}~\bibnamefont {Sato}}, \bibinfo {author}
  {\bibfnamefont {F.}~\bibnamefont {Matsukura}},\ and\ \bibinfo {author}
  {\bibfnamefont {H.}~\bibnamefont {Ohno}},\ }\bibfield  {title} {\enquote
  {\bibinfo {title} {{Magnetic-field-angle dependence of coercivity in
  CoFeB/MgO magnetic tunnel junctions with perpendicular easy axis}},}\
  }\href@noop {} {\bibfield  {journal} {\bibinfo  {journal} {Applied Physics
  Letters}\ }\textbf {\bibinfo {volume} {111}},\ \bibinfo {pages} {132407}
  (\bibinfo {year} {2017})}\BibitemShut {NoStop}%
\bibitem [{\citenamefont {Moutafis}\ \emph {et~al.}(2007)\citenamefont
  {Moutafis}, \citenamefont {Komineas}, \citenamefont {Vaz}, \citenamefont
  {Bland}, \citenamefont {Shima}, \citenamefont {Seki},\ and\ \citenamefont
  {Takanashi}}]{Moutafis2007}%
  \BibitemOpen
  \bibfield  {author} {\bibinfo {author} {\bibfnamefont {C.}~\bibnamefont
  {Moutafis}}, \bibinfo {author} {\bibfnamefont {S.}~\bibnamefont {Komineas}},
  \bibinfo {author} {\bibfnamefont {C.~A.}\ \bibnamefont {Vaz}}, \bibinfo
  {author} {\bibfnamefont {J.~A.}\ \bibnamefont {Bland}}, \bibinfo {author}
  {\bibfnamefont {T.}~\bibnamefont {Shima}}, \bibinfo {author} {\bibfnamefont
  {T.}~\bibnamefont {Seki}},\ and\ \bibinfo {author} {\bibfnamefont
  {K.}~\bibnamefont {Takanashi}},\ }\bibfield  {title} {\enquote {\bibinfo
  {title} {{Magnetic bubbles in FePt nanodots with perpendicular
  anisotropy}},}\ }\href@noop {} {\bibfield  {journal} {\bibinfo  {journal}
  {Physical Review B}\ }\textbf {\bibinfo {volume} {76}},\ \bibinfo {pages}
  {104426} (\bibinfo {year} {2007})}\BibitemShut {NoStop}%
\bibitem [{\citenamefont {Lavrijsen}\ \emph {et~al.}(2010)\citenamefont
  {Lavrijsen}, \citenamefont {Franken}, \citenamefont {Kohlhepp}, \citenamefont
  {Swagten},\ and\ \citenamefont {Koopmans}}]{Lavrijsen2010}%
  \BibitemOpen
  \bibfield  {author} {\bibinfo {author} {\bibfnamefont {R.}~\bibnamefont
  {Lavrijsen}}, \bibinfo {author} {\bibfnamefont {J.~H.}\ \bibnamefont
  {Franken}}, \bibinfo {author} {\bibfnamefont {J.~T.}\ \bibnamefont
  {Kohlhepp}}, \bibinfo {author} {\bibfnamefont {H.~J.}\ \bibnamefont
  {Swagten}},\ and\ \bibinfo {author} {\bibfnamefont {B.}~\bibnamefont
  {Koopmans}},\ }\bibfield  {title} {\enquote {\bibinfo {title} {{Controlled
  domain-wall injection in perpendicularly magnetized strips}},}\ }\href@noop
  {} {\bibfield  {journal} {\bibinfo  {journal} {Applied Physics Letters}\
  }\textbf {\bibinfo {volume} {96}},\ \bibinfo {pages} {222502} (\bibinfo
  {year} {2010})}\BibitemShut {NoStop}%
\bibitem [{\citenamefont {Talapatra}\ and\ \citenamefont
  {Mohanty}(2016)}]{Talapatra2016}%
  \BibitemOpen
  \bibfield  {author} {\bibinfo {author} {\bibfnamefont {A.}~\bibnamefont
  {Talapatra}}\ and\ \bibinfo {author} {\bibfnamefont {J.}~\bibnamefont
  {Mohanty}},\ }\bibfield  {title} {\enquote {\bibinfo {title} {{Laser induced
  local modification of magnetic domain in Co/Pt multilayer}},}\ }\href@noop {}
  {\bibfield  {journal} {\bibinfo  {journal} {Journal of Magnetism and Magnetic
  Materials}\ }\textbf {\bibinfo {volume} {418}},\ \bibinfo {pages} {224--230}
  (\bibinfo {year} {2016})}\BibitemShut {NoStop}%
\bibitem [{\citenamefont {Shepley}\ \emph {et~al.}(2018)\citenamefont
  {Shepley}, \citenamefont {Tunnicliffe}, \citenamefont {Shahbazi},
  \citenamefont {Burnell},\ and\ \citenamefont {Moore}}]{Shepley2018}%
  \BibitemOpen
  \bibfield  {author} {\bibinfo {author} {\bibfnamefont {P.~M.}\ \bibnamefont
  {Shepley}}, \bibinfo {author} {\bibfnamefont {H.}~\bibnamefont
  {Tunnicliffe}}, \bibinfo {author} {\bibfnamefont {K.}~\bibnamefont
  {Shahbazi}}, \bibinfo {author} {\bibfnamefont {G.}~\bibnamefont {Burnell}},\
  and\ \bibinfo {author} {\bibfnamefont {T.~A.}\ \bibnamefont {Moore}},\
  }\bibfield  {title} {\enquote {\bibinfo {title} {{Magnetic properties,
  domain-wall creep motion, and the Dzyaloshinskii-Moriya interaction in
  Pt/Co/Ir thin films}},}\ }\href@noop {} {\bibfield  {journal} {\bibinfo
  {journal} {Physical Review B}\ }\textbf {\bibinfo {volume} {97}},\ \bibinfo
  {pages} {134417} (\bibinfo {year} {2018})}\BibitemShut {NoStop}%
\bibitem [{\citenamefont {Mangin}\ \emph {et~al.}(2006)\citenamefont {Mangin},
  \citenamefont {Ravelosona}, \citenamefont {Katine}, \citenamefont {Carey},
  \citenamefont {Terris},\ and\ \citenamefont {Fullerton}}]{Mangin2006}%
  \BibitemOpen
  \bibfield  {author} {\bibinfo {author} {\bibfnamefont {S.}~\bibnamefont
  {Mangin}}, \bibinfo {author} {\bibfnamefont {D.}~\bibnamefont {Ravelosona}},
  \bibinfo {author} {\bibfnamefont {J.~A.}\ \bibnamefont {Katine}}, \bibinfo
  {author} {\bibfnamefont {M.~J.}\ \bibnamefont {Carey}}, \bibinfo {author}
  {\bibfnamefont {B.~D.}\ \bibnamefont {Terris}},\ and\ \bibinfo {author}
  {\bibfnamefont {E.~E.}\ \bibnamefont {Fullerton}},\ }\bibfield  {title}
  {\enquote {\bibinfo {title} {{Current-induced magnetization reversal in
  nanopillars with perpendicular anisotropy}},}\ }\href@noop {} {\bibfield
  {journal} {\bibinfo  {journal} {Nature Materials}\ }\textbf {\bibinfo
  {volume} {5}},\ \bibinfo {pages} {210--215} (\bibinfo {year}
  {2006})}\BibitemShut {NoStop}%
\bibitem [{\citenamefont {Williams}(1982)}]{Williams1982}%
  \BibitemOpen
  \bibfield  {author} {\bibinfo {author} {\bibfnamefont {E.~M.}\ \bibnamefont
  {Williams}},\ }\bibfield  {title} {\enquote {\bibinfo {title} {{The Dorf
  Effect: Magnetization Ripple In Particulate Media}},}\ }\href@noop {}
  {\bibfield  {journal} {\bibinfo  {journal} {IEEE Transactions on Magnetics}\
  }\textbf {\bibinfo {volume} {MAG-18}},\ \bibinfo {pages} {1086} (\bibinfo
  {year} {1982})}\BibitemShut {NoStop}%
\bibitem [{\citenamefont {Proksch}\ \emph {et~al.}(1995)\citenamefont
  {Proksch}, \citenamefont {Runge}, \citenamefont {Hansma}, \citenamefont
  {Foss},\ and\ \citenamefont {Walsh}}]{Proksch1995}%
  \BibitemOpen
  \bibfield  {author} {\bibinfo {author} {\bibfnamefont {R.}~\bibnamefont
  {Proksch}}, \bibinfo {author} {\bibfnamefont {E.}~\bibnamefont {Runge}},
  \bibinfo {author} {\bibfnamefont {P.~K.}\ \bibnamefont {Hansma}}, \bibinfo
  {author} {\bibfnamefont {S.}~\bibnamefont {Foss}},\ and\ \bibinfo {author}
  {\bibfnamefont {B.}~\bibnamefont {Walsh}},\ }\bibfield  {title} {\enquote
  {\bibinfo {title} {{High field magnetic force microscopy}},}\ }\href@noop {}
  {\bibfield  {journal} {\bibinfo  {journal} {Journal of Applied Physics}\
  }\textbf {\bibinfo {volume} {78}},\ \bibinfo {pages} {3303--3307} (\bibinfo
  {year} {1995})}\BibitemShut {NoStop}%
\bibitem [{\citenamefont {Gomez}, \citenamefont {Burke},\ and\ \citenamefont
  {Mayergoyz}(1996)}]{Gomez1996}%
  \BibitemOpen
  \bibfield  {author} {\bibinfo {author} {\bibfnamefont {R.~D.}\ \bibnamefont
  {Gomez}}, \bibinfo {author} {\bibfnamefont {E.~R.}\ \bibnamefont {Burke}},\
  and\ \bibinfo {author} {\bibfnamefont {I.~D.}\ \bibnamefont {Mayergoyz}},\
  }\bibfield  {title} {\enquote {\bibinfo {title} {{Magnetic imaging in the
  presence of external fields: Technique and applications (invited)}},}\ }\href
  {https://doi.org/10.1063/1.361966} {\bibfield  {journal} {\bibinfo  {journal}
  {Journal of Applied Physics}\ }\textbf {\bibinfo {volume} {79}},\ \bibinfo
  {pages} {6441} (\bibinfo {year} {1996})}\BibitemShut {NoStop}%
\bibitem [{\citenamefont {Mohanty}, \citenamefont {Engel-Herbert},\ and\
  \citenamefont {Hesjedal}(2005)}]{Mohanty2005}%
  \BibitemOpen
  \bibfield  {author} {\bibinfo {author} {\bibfnamefont {J.}~\bibnamefont
  {Mohanty}}, \bibinfo {author} {\bibfnamefont {R.}~\bibnamefont
  {Engel-Herbert}},\ and\ \bibinfo {author} {\bibfnamefont {T.}~\bibnamefont
  {Hesjedal}},\ }\bibfield  {title} {\enquote {\bibinfo {title} {{Variable
  magnetic field and temperature magnetic force microscopy}},}\ }\href@noop {}
  {\bibfield  {journal} {\bibinfo  {journal} {Applied Physics A}\ }\textbf
  {\bibinfo {volume} {81}},\ \bibinfo {pages} {1359} (\bibinfo {year}
  {2005})}\BibitemShut {NoStop}%
\bibitem [{\citenamefont {Harroun}\ \emph {et~al.}(2006)\citenamefont
  {Harroun}, \citenamefont {Desrochers}, \citenamefont {Nieh}, \citenamefont
  {Watson},\ and\ \citenamefont {Katsaras}}]{Harroun2006}%
  \BibitemOpen
  \bibfield  {author} {\bibinfo {author} {\bibfnamefont {T.~A.}\ \bibnamefont
  {Harroun}}, \bibinfo {author} {\bibfnamefont {C.~M.}\ \bibnamefont
  {Desrochers}}, \bibinfo {author} {\bibfnamefont {M.~P.}\ \bibnamefont
  {Nieh}}, \bibinfo {author} {\bibfnamefont {M.~J.}\ \bibnamefont {Watson}},\
  and\ \bibinfo {author} {\bibfnamefont {J.}~\bibnamefont {Katsaras}},\
  }\bibfield  {title} {\enquote {\bibinfo {title} {{0.9 T static magnetic field
  and temperature-controlled specimen environment for use with general-purpose
  optical microscopes}},}\ }\href {https://doi.org/10.1063/1.2162433}
  {\bibfield  {journal} {\bibinfo  {journal} {Review of Scientific
  Instruments}\ }\textbf {\bibinfo {volume} {77}},\ \bibinfo {pages} {014102}
  (\bibinfo {year} {2006})}\BibitemShut {NoStop}%
\bibitem [{\citenamefont {Oldenbourg}\ and\ \citenamefont
  {Phillips}(1986)}]{Oldenbourg1986}%
  \BibitemOpen
  \bibfield  {author} {\bibinfo {author} {\bibfnamefont {R.}~\bibnamefont
  {Oldenbourg}}\ and\ \bibinfo {author} {\bibfnamefont {W.~C.}\ \bibnamefont
  {Phillips}},\ }\bibfield  {title} {\enquote {\bibinfo {title} {{Small
  permanent magnet for fields up to 2.6 T}},}\ }\href
  {https://doi.org/10.1063/1.1138680} {\bibfield  {journal} {\bibinfo
  {journal} {Review of Scientific Instruments}\ }\textbf {\bibinfo {volume}
  {57}},\ \bibinfo {pages} {2362--2365} (\bibinfo {year} {1986})}\BibitemShut
  {NoStop}%
\bibitem [{\citenamefont {Cugat}, \citenamefont {Hansson},\ and\ \citenamefont
  {Coey}(1994)}]{Cugat1994}%
  \BibitemOpen
  \bibfield  {author} {\bibinfo {author} {\bibfnamefont {O.}~\bibnamefont
  {Cugat}}, \bibinfo {author} {\bibfnamefont {P.}~\bibnamefont {Hansson}},\
  and\ \bibinfo {author} {\bibfnamefont {J.~M.~D.}\ \bibnamefont {Coey}},\
  }\bibfield  {title} {\enquote {\bibinfo {title} {{Permanent Magnet Variable
  Flux Sources}},}\ }\href {https://doi.org/10.1109/20.334162} {\bibfield
  {journal} {\bibinfo  {journal} {IEEE Transactions on Magnetics}\ }\textbf
  {\bibinfo {volume} {30}},\ \bibinfo {pages} {4602--4604} (\bibinfo {year}
  {1994})}\BibitemShut {NoStop}%
\bibitem [{\citenamefont {Duong}\ \emph {et~al.}(2019)\citenamefont {Duong},
  \citenamefont {Raju}, \citenamefont {Petrovi{\'{c}}}, \citenamefont
  {Tomasello}, \citenamefont {Finocchio},\ and\ \citenamefont
  {Panagopoulos}}]{Duong2019}%
  \BibitemOpen
  \bibfield  {author} {\bibinfo {author} {\bibfnamefont {N.~K.}\ \bibnamefont
  {Duong}}, \bibinfo {author} {\bibfnamefont {M.}~\bibnamefont {Raju}},
  \bibinfo {author} {\bibfnamefont {A.~P.}\ \bibnamefont {Petrovi{\'{c}}}},
  \bibinfo {author} {\bibfnamefont {R.}~\bibnamefont {Tomasello}}, \bibinfo
  {author} {\bibfnamefont {G.}~\bibnamefont {Finocchio}},\ and\ \bibinfo
  {author} {\bibfnamefont {C.}~\bibnamefont {Panagopoulos}},\ }\bibfield
  {title} {\enquote {\bibinfo {title} {{Stabilizing zero-field skyrmions in
  Ir/Fe/Co/Pt thin film multilayers by magnetic history control}},}\ }\href
  {https://doi.org/10.1063/1.5080713} {\bibfield  {journal} {\bibinfo
  {journal} {Appl. Phys. Lett.}\ }\textbf {\bibinfo {volume} {114}},\ \bibinfo
  {pages} {072401} (\bibinfo {year} {2019})}\BibitemShut {NoStop}%
\bibitem [{Asy()}]{Asylum}%
  \BibitemOpen
  \href@noop {} {}\bibinfo {note} {Model VFM4, Asylum Research, Oxford
  Instruments, Goleta, CA}\BibitemShut {NoStop}%
\bibitem [{NTE()}]{NTEGRA}%
  \BibitemOpen
  \href@noop {} {}\bibinfo {note} {Vertical field option, NTEGRA
  nanolaboratory, NT-MDT Spectrum Instruments, Moscow, Russia}\BibitemShut
  {NoStop}%
\bibitem [{AEv()}]{AEvico1}%
  \BibitemOpen
  \href@noop {} {}\bibinfo {note} {Perpendicular Electromagnet, Evico Magnetics
  GmbH, Dresden, Germany}\BibitemShut {NoStop}%
\bibitem [{BEv()}]{BEvico2}%
  \BibitemOpen
  \href@noop {} {}\bibinfo {note} {Perpendicular Coil, Evico Magnetics GmbH,
  Dresden, Germany}\BibitemShut {NoStop}%
\bibitem [{\citenamefont {Meeker}()}]{FEMM}%
  \BibitemOpen
  \bibfield  {author} {\bibinfo {author} {\bibfnamefont {D.}~\bibnamefont
  {Meeker}},\ }\href@noop {} {\enquote {\bibinfo {title} {{Finite Element
  Method Magnetics}},}\ }\bibinfo {note} {\url{www.femm.info}}\BibitemShut
  {NoStop}%
\bibitem [{AKe()}]{AKepcoBOP}%
  \BibitemOpen
  \href@noop {} {}\bibinfo {note} {Model BOP 36-28GL, Kepco Inc., Flushing,
  NY}\BibitemShut {NoStop}%
\bibitem [{Mas()}]{Masterbond}%
  \BibitemOpen
  \href@noop {} {}\bibinfo {note} {EP21TCHT-1 single-component epoxy, Master
  Bond Inc., Hackensack, NJ}\BibitemShut {NoStop}%
\bibitem [{Sup()}]{Superior}%
  \BibitemOpen
  \href@noop {} {}\bibinfo {note} {.060''x.175'' Heavy GP/MR-200 rectangular MW
  36 copper magnet wire, Superior Essex Inc., Atlanta, GA}\BibitemShut
  {NoStop}%
\bibitem [{Ela()}]{Elantas}%
  \BibitemOpen
  \href@noop {} {}\bibinfo {note} {MC4236 / W4236 two component epoxy potting
  compound, Elantas Malaysia Sdn. Bhd., Selangor Darul Ehsan,
  Malaysia}\BibitemShut {NoStop}%
\bibitem [{Hal()}]{Hall}%
  \BibitemOpen
  \href@noop {} {}\bibinfo {note} {Model XHGT-9060, Lake Shore Cryotronics
  Inc., Westerville, OH}\BibitemShut {NoStop}%
\bibitem [{Vac()}]{Vacoflux}%
  \BibitemOpen
  \href@noop {} {}\bibinfo {note} {Vacoflux 50, Vacuumschmelze GmbH, Hanau,
  Germany}\BibitemShut {NoStop}%
\bibitem [{ALa()}]{ALakeshorePt}%
  \BibitemOpen
  \href@noop {} {}\bibinfo {note} {Model PT-103, Lake Shore Cryotronics Inc.,
  Westerville, OH}\BibitemShut {NoStop}%
\bibitem [{BLa()}]{BLakeshore}%
  \BibitemOpen
  \href@noop {} {}\bibinfo {note} {Model 325, Lake Shore Cryotronics Inc.,
  Westerville, OH}\BibitemShut {NoStop}%
\bibitem [{Pel()}]{Peltier}%
  \BibitemOpen
  \href@noop {} {}\bibinfo {note} {Adaptive Models ET-063-08-15 and
  ET-031-08-15-RS, European Thermodynamics Ltd., Kibworth, United
  Kingdom}\BibitemShut {NoStop}%
\bibitem [{AEp()}]{AEpotek1}%
  \BibitemOpen
  \href@noop {} {}\bibinfo {note} {Epo-Tek E4110, Epoxy Technology Inc.,
  Billerica, MA}\BibitemShut {NoStop}%
\bibitem [{BKe()}]{BKepcoPTR}%
  \BibitemOpen
  \href@noop {} {}\bibinfo {note} {Model PTR21-2.5, Kepco Inc., Flushing,
  NY}\BibitemShut {NoStop}%
\bibitem [{Rog()}]{Rogers}%
  \BibitemOpen
  \href@noop {} {}\bibinfo {note} {Rogers Corporation RO3003, Chandler,
  AZ}\BibitemShut {NoStop}%
\bibitem [{BEp()}]{BEpotek2}%
  \BibitemOpen
  \href@noop {} {}\bibinfo {note} {Epo-Tek H70E-2, Epoxy Technology Inc.,
  Billerica, MA}\BibitemShut {NoStop}%
\bibitem [{HEx()}]{HExch}%
  \BibitemOpen
  \href@noop {} {}\bibinfo {note} {Models LI-102 and LI-301, Thermo Electric
  Devices, Draycott, United Kingdom}\BibitemShut {NoStop}%
\bibitem [{Jul()}]{Julabo}%
  \BibitemOpen
  \href@noop {} {}\bibinfo {note} {Model FL601, Julabo GmbH, Seelbach,
  Germany}\BibitemShut {NoStop}%
\end{thebibliography}
\end{document}